\documentclass[twocolumn]{revtex4}
\usepackage[paper=letterpaper,top=0.8in,bottom=0.8in,left=0.7in,right=0.7in]{geometry}

\usepackage{amsmath}
\usepackage{amsfonts}
\usepackage{amsthm}
\usepackage{amssymb}
\usepackage{graphicx}
\usepackage{mathrsfs}
\usepackage{latexsym}
\usepackage{graphics}
\usepackage{cases}
\usepackage{hyperref}
\usepackage{mathtools}
\usepackage{dsfont}

\renewcommand{\d}{{\rm d}}
\newcommand{\Tc}{T_{\rm c}}
\newcommand{\Rm}{R_0}
\newcommand{\pv}{p_{\rm v}}

\begin{document}

\title{Analytical Approximations for the Collapse of an Empty Spherical Bubble}

\author{D. Obreschkow$^1$}
\author{M. Bruderer$^2$}
\author{M. Farhat$^3$}

\affiliation{
\mbox{$^1\,$The University of Western Australia, ICRAR, 35 Stirling Hwy,
Crawley, WA 6009, Australia}
\mbox{$^2\,$Fachbereich Physik, Universit\"{a}t Konstanz, D-78457 Konstanz,
Germany}
\mbox{$^3\,$Ecole Polytechnique F\'{e}d\'{e}rale de Lausanne, LMH, 1007 Lausanne, Switzerland}\\
\\}


\date{\today}

\begin{abstract}
The Rayleigh equation $\frac{3}{2}\dot{R}+R\ddot{R}+p\rho^{-1}=0$ with initial conditions $R(0)=\Rm$, $\dot{R}(0)=0$ models the collapse of an empty spherical bubble of radius $R(T)$ in an ideal, infinite liquid with far-field pressure $p$ and density $\rho$. The solution for $r\equiv R/\Rm$ as a function of time $t\equiv T/\Tc$, where $R(\Tc)\equiv0$, is independent of $\Rm$, $p$, and $\rho$. While no closed-form expression for $r(t)$ is known we find that $r_0(t)=(1-t^2)^{2/5}$ approximates $r(t)$ with an error below 1\%. A systematic development in orders of $t^2$ further yields the 0.001\%--approximation $r_\ast(t)=r_0(t)[1-a_1\,{\rm Li}_{2.21}(t^2)]$, where $a_1\approx-0.01832099$ is a constant and Li is the polylogarithm. The usefulness of these approximations is demonstrated by comparison to high-precision cavitation data obtained in microgravity. 
\end{abstract}

\maketitle


\section{Introduction}
George Gabriel Stokes might not have anticipated the importance of his endeavor when challenging his students in 1847 to calculate the collapse motion of an empty bubble in water \citep{Stokes1847}. The reach of this academic exercise was recognized in 1917 by Lord Rayleigh \citep{Rayleigh1917} who conveyed a link between collapsing bubbles and the erosion damage found on ship propellers. The basic equation of motion of a collapsing bubble \citep{Brennen1995}, now known as the Rayleigh equation (RE), reads
\begin{equation}\label{eq_rayleigh}
	\frac{3}{2}\left(\frac{\d R}{\d T}\right)^2+\frac{\d^2 R}{\d T^2}R+k = 0,
\end{equation}
where the bubble radius $R$ is a function of the time $T$, and $k$ is a constant. Given the initial conditions
\begin{equation}\label{eq_conditions}
	R(0)=\Rm,~\left.\frac{\d R}{\d T}\right|_{T=0}=0
\end{equation}
and the definition $k\equiv p\rho^{-1}$, the RE describes the collapse of an empty spherical bubble of initial radius $\Rm$ in an incompressible, inviscid, infinite liquid with uniform far-field pressure $p$ and density $\rho$. If, alternatively, $k$ is defined as $(p-\pv)\rho^{-1}$, the RE extends to the case of a gas-filled bubble with constant inner pressure $\pv$. The RE neglects non-condensable bubble gas, as well as surface tension and viscosity \citep{Plesset1977}, liquid compressibility \citep{Prosperetti1987}, and thermal effects \citep{Dergarabedian1953}. However, regardless of those limitations and various enhanced models available today \citep{Plesset1949,Prosperetti1987,Dergarabedian1953,Plesset1971,Obreschkow2006,Obreschkow2011b}, the RE remains widely used in practice, owing to its simplicity and often sufficient accuracy. This is despite the fact that the RE yields no closed-form solution for $k>0$. While numerical solutions can be obtained, systematic analytical approximations offer better insight into the mathematical nature of the collapse, as we will show in this paper. Analytical approximations also become handy when the RE is integrated into multi-scale models, and they offer an intuitive understanding for the collapse motion.

In this paper we develop highly accurate, yet remarkably efficient analytical approximations for the solution of the RE. We first recall the standard normalization of the RE, based on which the analytical approximations are then obtained in a systematic way. These approximations are then compared against high-precision measurements of the most spherical bubbles available today. A short discussion concludes the paper.



\section{Normalized Rayleigh Model}\label{section_normalization}
Multypling Eq.~(\ref{eq_rayleigh}) by $3R^2(\d R/\d T)$, then integrating with respect to $T$, and expressing the integration constant using the initial conditions in Eq.~(\ref{eq_conditions}), we find
\begin{equation}\label{eq_conservation}
	\frac{3}{2}\left(\frac{\d R}{\d T}\right)^2R^3+kR^3 = k\Rm^3. 
\end{equation}
Up to a factor $4\pi\rho/3$, Eq.~(\ref{eq_conservation}) expresses the conservation of energy. Note that Eq.~(\ref{eq_conservation}) only implies Eq.~(\ref{eq_rayleigh}) if $R^2(\d R/\d T)\neq0$, and according to the initial conditions in Eq.~(\ref{eq_conditions}) this relation breaks down at $T=0$. Nevertheless, we can still refer to Eq.~(\ref{eq_conservation}) to analyze $R(T)$ for $T>0$. In particular, the collapse time $\Tc$, defined by $R(\Tc)\equiv0$, is found by integrating $\d T$ from $0$ to $\Tc$ and $\d R$ from $\Rm$ to 0. We obtain
\begin{equation}\label{eq_collpase_time}
	\Tc = \xi\,\Rm\,k^{-1/2},
\end{equation}
where $\xi\equiv\sqrt{3/2}\,\int_0^1(r^{-3}-1)^{-1/2}{\rm d}r\approx0.914681$ is a universal constant called the Rayleigh factor. Normalizing the radius to $r\equiv R/\Rm$ and the time to $t\equiv T/\Tc$, Eqs.~(\ref{eq_rayleigh}) and (\ref{eq_conditions}) become
%
%
\begin{equation}\label{eq_rayleigh_normalized}
	\frac{3}{2}\dot{r}^2+\ddot{r}r+\xi^2 = 0,
\end{equation}
\vspace{-0.5cm}
\begin{equation}\label{eq_conditions_normalized}
	r(0)=1,~\dot{r}(0)=0,
\end{equation}
where dots denote derivatives with respect to $t$. Using the same normalization, Eq.~(\ref{eq_conservation}) translates to
\begin{equation}\label{eq_conservation_normalized}
	\dot{r}^2 = \frac{2}{3}\xi^2\big(r^{-3}-1\big).
\end{equation}
Substituting Eq.~(\ref{eq_conservation_normalized}) back into Eq.~(\ref{eq_rayleigh_normalized}) then implies
\begin{equation}\label{eq_rayleigh_alternative}
	\ddot{r} = -\xi^2r^{-4},
\end{equation}
which is an interesting, but little known alternative form of the normalized RE given in Eq.~(\ref{eq_rayleigh_normalized}).

The key advantage of Eq.~(\ref{eq_rayleigh_normalized}) over Eq.~(\ref{eq_rayleigh}) is its invariance with respect to $\Rm$ and $k$. Stated differently, we only need to solve Eq.~(\ref{eq_rayleigh_normalized}) once in order to solve Eq.~(\ref{eq_rayleigh}) for any choice of $\Rm>0$ and $k>0$. The solution of Eq.~(\ref{eq_rayleigh_normalized}) in the range $t=[0,1]$ is displayed in Fig.~\ref{fig_models}(b). This solution was obtained using a Cash-Karp fourth-fifth order Runge-Kutta method \citep{Cash1990}. The relative error made on the collapse time lies below $10^{-15}$, and thus our numerical solution for $r(t)$ can be considered exact as far as this article is concerned.


\section{Analytical approximations}\label{section_approximations}
\begin{figure}
	\includegraphics[width=83mm]{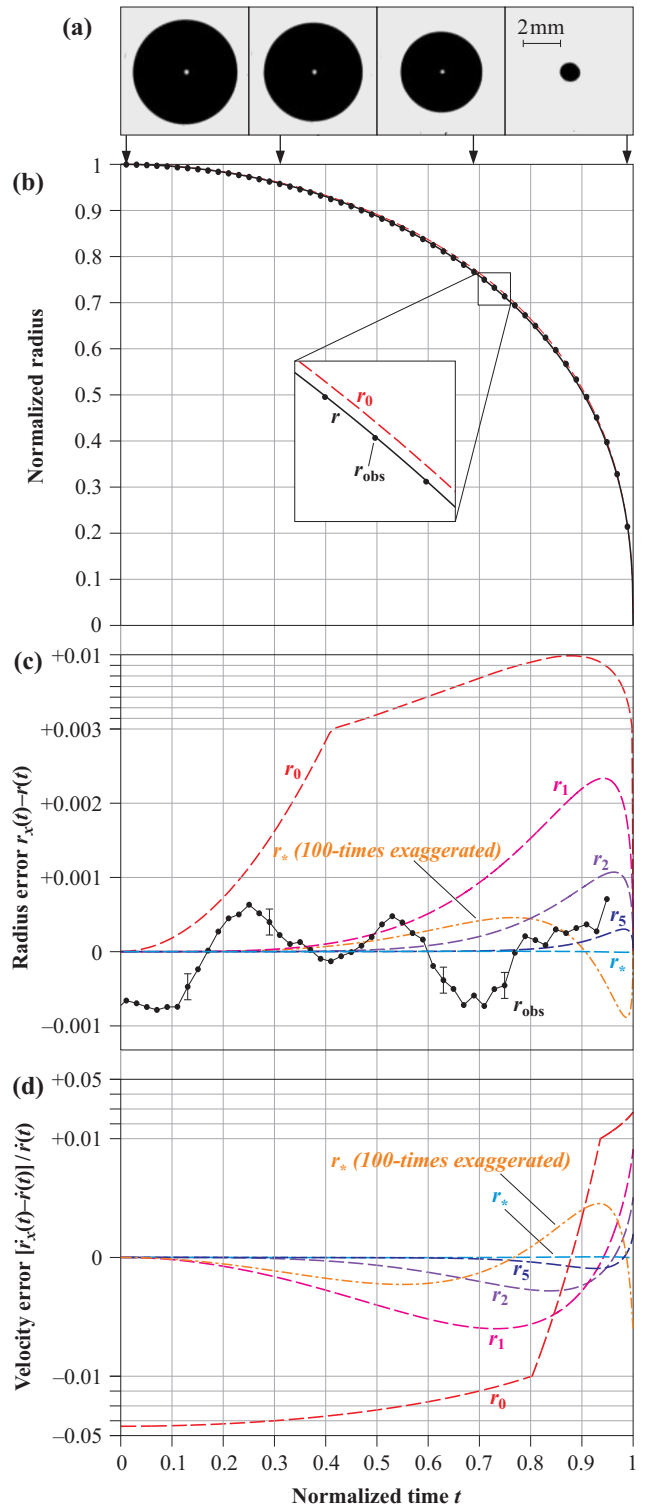}
	\caption{(Color online) (a) Four subsequent high-speed images of the collapsing spherical bubble in microgravity. (b) Collapse functions: exact solution $r(t)$ of Eqs.~(\ref{eq_rayleigh_normalized}) and (\ref{eq_conditions_normalized}) (solid line), measurement $r_{\rm obs}(t)$ (dots), and zeroth-order analytical approximation $r_0(t)$ (dashed line). (c) Errors of $r_{\rm obs}(t)$ and the analytical models $r_n(t)$ [Eq.~\ref{eq_approx_n}] and $r_\ast(t)$ [Eq.~\ref{eq_approx_star}] relative to the exact solution $r(t)$. Bars represent 67\% statistical measurement uncertainties. (d) Errors of the velocities $\dot{r}_{\rm obs}(t)$, $\dot{r}_\ast(t)$, and $\dot{r}_n(t)$ relative to $\dot{r}(t)$.}
	\label{fig_models}
\end{figure}

To find analytical approximations for $r(t)$ we first notice that the Rayleigh model given in~Eqs.~(\ref{eq_rayleigh_normalized}) and (\ref{eq_conditions_normalized}) is symmetric in time. It follows that $r(t)=r(-t)$ can necessarily be expressed as a function of $t^2$. Second, we observe that $r(t)$ is non--analytic at $t=\pm1$, as can be seen from the divergence of $\dot{r}$ as $r\rightarrow0$ in Eq.~(\ref{eq_conservation_normalized}). Since $r(t)$ converges as $t\rightarrow\pm1$, the singularities are neither essential singularities nor poles, and therefore must be branch points induced by the double-valued square root appearing when Eq.~(\ref{eq_conservation_normalized}) is solved for $\dot{r}$. The simplest function exhibiting such a pair of branch points at $t=\pm1$ is the power law $r_0(t)\equiv(1-t^2)^\alpha$ with $\alpha\in\,]0,1[$. In order to use $r_0(t)$ as an approximation of $r(t)$ the parameter $\alpha$ can be determined in two ways. First, we can request that $\ddot{r}_0(0)=\ddot{r}(0)$, which together with Eqs.~(\ref{eq_conditions_normalized}) and (\ref{eq_rayleigh_alternative}) implies $\alpha=\xi^2/2\approx0.418321$. Alternatively, we can impose that $\dot{r}_0(t)$ and $\dot{r}(t)$ exhibit similar asymptotic behavior at $t=1$. In fact, Eq.~(\ref{eq_approx_0}) implies $\dot{r}_0\propto r_0^{1-1/\alpha}$ and Eq.~(\ref{eq_conservation_normalized}) implies $\dot{r}\propto r^{-3/2}$ as $t\rightarrow1$. Matching the powers of the two asymptotic functions we find $\alpha=2/5=0.4$. The coincidental close similarity of $\alpha$ determined at $t=0$ and $t=1$ suggests that $r_0(t)$ is a good approximation of $r(t)$ for both values of $\alpha$. Here we choose $\alpha=2/5$ and the corresponding approximation
\begin{equation}\label{eq_approx_0}
	r_0(t)\equiv(1-t^2)^{^\frac{2}{5}},
\end{equation}
hence accepting that $\ddot{r}_0(0)\neq\ddot{r}(0)$ for the moment. This approximation is displayed in Fig.~\ref{fig_models}(b). Its similarity to the Rayleigh function $r(t)$ is demonstrated by the small residual $r_0(t)-r(t)$, shown in Fig.~\ref{fig_models}(c). Indeed, $r_0(t)$ never differs by more than $0.01$ from $r(t)$, and by construction we have $r_0(t)=r(t)$ at $t\in\{-1,0,1\}$. The normalized velocity residual $[\dot{r}_0(t)-\dot{r}(t)]/\dot{r}(t)$, shown in Fig.~\ref{fig_models}(d), takes absolute values up to $0.044$. The fact that $\lim_{t\rightarrow0}[\dot{r}_0(t)-\dot{r}(t)]/\dot{r}(t)=\ddot{r}_0(0)/\ddot{r}(0)-1$ differs from zero explicitly reveals that $\ddot{r}_0(0)\neq\ddot{r}(0)$.

\begin{table}[t!]
   \begin{tabular}{cccccc}
     \hline
     ~$q$~ & $a_q$ & $\!\!\!\!\log_{10}\sigma(r_q)$ & $\log_{10}\epsilon(r_q)$ & $\log_{10}\dot{\sigma}(r_q)$ & $\log_{10}\dot{\epsilon}(r_q)$ \\
     \hline
$ 0$ & $1$ & $-2.5$ & $-2.0$ & $-1.7$ & $-1.4$ \\
$ 1$ & $-0.01832099$ & $-3.1$ & $-2.6$ & $-2.6$ & $-2.0$ \\
$ 2$ & $-0.00399003$ & $-3.5$ & $-3.0$ & $-2.9$ & $-2.3$ \\
$ 3$ & $-0.00161041$ & $-3.8$ & $-3.2$ & $-3.2$ & $-2.5$ \\
$ 4$ & $-0.00084483$ & $-4.0$ & $-3.4$ & $-3.4$ & $-2.6$ \\
$ 5$ & $-0.00051245$ & $-4.2$ & $-3.5$ & $-3.5$ & $-2.7$ \\
$ 6$ & $-0.00034081$ & $-4.3$ & $-3.6$ & $-3.6$ & $-2.8$ \\
$ 7$ & $-0.00024153$ & $-4.4$ & $-3.7$ & $-3.7$ & $-2.9$ \\
$ 8$ & $-0.00017930$ & $-4.5$ & $-3.8$ & $-3.8$ & $-2.9$ \\
$ 9$ & $-0.00013791$ & $-4.6$ & $-3.9$ & $-3.9$ & $-3.0$ \\
$10$ & $-0.00010908$ & $-4.7$ & $-4.0$ & $-4.0$ & $-3.1$ \\
	 \hline
\multicolumn{2}{c}{Function $r_\ast(t)$} & $-5.6$ & $-5.1$ & $-4.7$ & $-4.2$ \\
     \hline
   \end{tabular}
   \caption{Coefficients $a_q$ in the summations of Eqs.~(\ref{eq_approx_inf}) and (\ref{eq_approx_n}), and accuracies of the approximations $r_q(t)$.}
   \label{tab_aq}
\end{table}

\begin{figure}
	\includegraphics[width=83mm]{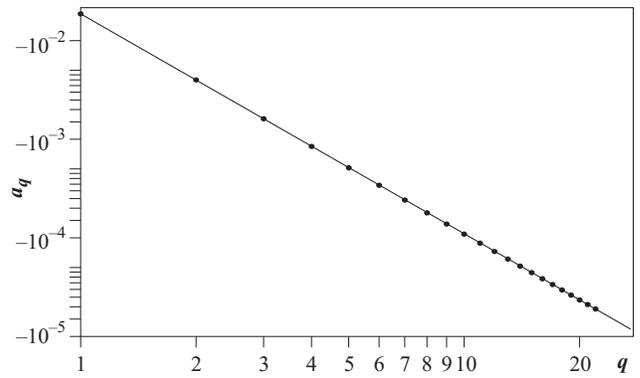}
	\caption{Dots: analytically calculated coefficients $a_q$ in the expansion of Eq.~(\ref{eq_approx_inf}). Solid line: power law fit of Eq.~(\ref{eq_approx_aq}), which has been forced to pass through $a_1$.}
	\label{fig_aq}
\end{figure}

We now improve the accuracy of our approximation $r_0(t)$ at $t=0$ through the modified ansatz $r_\infty(t)\equiv r_0(t)f(t)$, where $f(t)$ is a smooth function defined by the condition that $\d^{q}r_\infty(0)/\d t^{q}=\d^{q}r(0)/\d t^{q}$ for all derivatives of order $q\geq1$. As we will see, this infinite set of boundary conditions can be met by restricting $f(t)$ to functions that can be expressed as Taylor series on the closed interval $t\in[-1,1]$. To respect the time--symmetry, $f(t)$ must be even, i.e.,~all odd powers in the Taylor series vanish. Thus,
\begin{equation}\label{eq_approx_inf}
	r_\infty(t) \equiv \big(1-t^2\big)^\frac{2}{5}\sum_{q=0}^\infty a_q t^{2q},
\end{equation}
where $a_q$ are real constants. The condition $r_\infty(0)=r(0)=1$ immediately implies $a_0=1$. All other coefficients $a_q$ are obtained by matching the even-order derivatives of  $r_\infty(t)$ and $r(t)$ at $t=0$. Those derivatives are evaluated analytically by differentiating Eqs.~(\ref{eq_approx_inf}) and (\ref{eq_rayleigh_alternative}), respectively, and applying the initial conditions given in Eq.~(\ref{eq_conditions_normalized}). We can proceed iteratively: first use $\d^{2q}r_\infty(0)/\d t^{2q}=\d^{2q}r(0)/\d t^{2q}$ with $q=1$ to get $a_1=2/5-\xi^2/2\approx-0.01832099$, then with $q=2$ to get $a_2=3/25+2a_1/5-\xi^4/6\approx-0.00399003$, and so forth. Note that all odd-order derivatives, such as $\dot{r}(0)$ and $\dot{r}_\infty(0)$, vanish due to time--symmetry. The analytical expressions for even-order derivatives of $r(t)$ and $r_\infty(t)$ get cumbersome as $q$ increases; however, the coefficients $a_q$ are easily obtained by using analytical software tools. The numerical values of $a_q$ up to $q=10$ are given in Table~\ref{tab_aq} and those up to $q=22$ are plotted in Fig.~\ref{fig_aq}. Using these values of the coefficients $a_q$ we can now construct approximations
\begin{equation}\label{eq_approx_n}
	r_n(t) \equiv \big(1-t^2\big)^{\frac{2}{5}}\sum_{q=0}^n a_q t^{2q}.
\end{equation}
of any order $n$. Since $a_q<0$ and $|a_q|<|a_{q-1}|$ for all $q>0$, the approximations $r_n(t)$ converge monotonically towards $r_\infty(t)$ as $n\rightarrow\infty$. Figures~\ref{fig_models}(c) and \ref{fig_models}(d) show that $r_n(t)$ and $\dot{r}_n(t)$ numerically converge towards the Rayleigh solution $r(t)$ and $\dot{r}(t)$. Already the second-order approximation $r_2(t)$ is roughly 10 times better than the zeroth-order approximation $r_0(t)$ discussed before. Yet, from a mathematical point of view, the crucial question is whether or not $r_\infty(t)$ is identical to $r(t)$ for all $t\in[-1,1]$, i.e.~if $r_\infty(t)$ is the solution of the normalized RE. The answer is no, since $\ddot{r}(t)$ in Eq.~(\ref{eq_rayleigh_alternative}) and $\ddot{r}_\infty(t)$ derived from Eq.~(\ref{eq_approx_inf}) do not obey the same asymptotic behavior as $t\rightarrow1$. In other words, $r_\infty(t)$ remains an approximation of $r(t)$, no matter the choice of the real coefficients $a_q$.

Figure~\ref{fig_aq} uncovers that $q$ and $a_q$ exhibit a remarkably tight power-law relation. Although this relation is not analytically exact, we can approximate the values of $a_q$ for $q>0$ as
\begin{equation}\label{eq_approx_aq}
	a_q \approx a_1\,q^{-2.21},
\end{equation}
with the exact $a_1=2/5-\xi^2/2\approx-0.01832099$ given in Table~\ref{tab_aq}. The relation given in Eq.~(\ref{eq_approx_aq}) is plotted as a solid line in Fig.~\ref{fig_aq}. Upon approximating the coefficients $a_q$ in $r_\infty(t)$ for $q>0$ by Eq.~(\ref{eq_approx_aq}) we obtain
\begin{equation}\label{eq_approx_star}
	r_\ast(t) \equiv \big(1-t^2\big)^{\frac{2}{5}}\Big[1+a_1\,{\rm Li}_{2.21}\big(t^2\big)\Big],
\end{equation}
where ${\rm Li}_s(x)\equiv\sum_{q=1}^\infty q^{-s}x^q$ is the polylogarithm, also known as Jonqui{\`e}re's function. Many programming languages contain ${\rm Li}_s(x)$ in their standard libraries. We emphasize that $r_\ast(t)$ slightly differs from $r_\infty(t)$ since the latter uses the exact coefficients $a_q$ rather than those approximated by Eq.~(\ref{eq_approx_aq}). Nonetheless, $r_\ast(t)$ is a very precise approximation of the Rayleigh solution $r(t)$ as can be seen from the residuals (multiplied by a factor 100) shown in Figs.~\ref{fig_models}(c) and \ref{fig_models}(d).

To quantify the accuracy of our approximations in a more refined way we consider the measures
\begin{eqnarray}
	\epsilon(r_x) & \equiv & \max_{t\in[0,1]}\big|r_x(t)-r(t)\big|\,, \label{eq_qual1} \\
	\sigma(r_x) & \equiv & \sqrt{\int_0^1\Big(r_x(t)-r(t)\Big)^2\d t}~, \label{eq_qual2} \\
	\dot{\epsilon}(r_x) & \equiv & \max_{t\in[0,1]}\left|\frac{\dot{r}_x(t)-\dot{r}(t)}{\dot{r}(t)}\right|\,, \label{eq_qual3} \\
	\dot{\sigma}(r_x) & \equiv & \sqrt{\int_0^1\left(\frac{\dot{r}_x(t)-\dot{r}(t)}{\dot{r}(t)}\right)^2\d t}~. \label{eq_qual4}
\end{eqnarray}
Here, $\epsilon(r_x)$ is the maximal error and $\sigma(r_x)$ the standard deviation of the approximation $r_x(t)$ with respect to the exact solution $r(t)$, and likewise $\dot{\epsilon}(r_x)$ and $\dot{\sigma}(r_x)$ give the accuracy of the derivative $\dot{r}_x(t)$. The residuals $\dot{r}_x(t)-\dot{r}(t)$ in Eqs.~(\ref{eq_qual3}) and (\ref{eq_qual4}) are normalized relative to $\dot{r}(t)$ to obtain meaningful, converging measures as $|\dot{r}|\rightarrow\infty$. It turns out that for all our approximations except $r_0(t)$ the absolute value of $[\dot{r}_x(t)-\dot{r}(t)]/\dot{r}(t)$ is maximal as $t\rightarrow1$. In this limit, the numerical evaluation of  $\dot{\epsilon}$ is delicate because of the divergence of $\dot{r}_x(t)$ and $\dot{r}(t)$. However, we can use the analytical expression $\lim_{t\rightarrow1}=[\dot{r}_x(t)-\dot{r}(t)]/\dot{r}(t)=(\sqrt{6}\alpha/\xi)^\alpha\,f(1)-1$, where $f(t)\equiv r_x(t)/(1-t^2)^{2/5}$ is the sum on the right-hand sides of Eqs.~(\ref{eq_approx_inf}), (\ref{eq_approx_n}), and (\ref{eq_approx_star}), respectively.

The logarithms of $\epsilon$, $\sigma$, $\dot{\epsilon}$, and $\dot{\sigma}$ for various approximations are given in Table~\ref{tab_aq}. These values support and extend the previous discussion of the residuals in Figs.~\ref{fig_models}(c) and \ref{fig_models}(d). In particular, we find that $r_\ast(t)$ approximates $r(t)$ with an error below $10^{-5}$ on the whole interval $t\in[0,1]$, while $\dot{r}_\ast(t)$ approximates $\dot{r}(t)$ at a relative error below than $10^{-4}$.


\section{Comparison to observed data}\label{section_application}
We now present a state-of-the-art experiment~(see details in Ref.~\citep{Obreschkow2011b}) of millimeter-sized bubbles with almost perfect sphericity. The high validity of the Rayleigh model for these bubbles  \citep{Lauterborn1972} stresses the need for accurate approximations such as $r_\ast(t)$ when analyzing such bubbles.

In the experiment, single bubbles grow inside a liquid from a point=-plasma generated by a mirror-focused nanosecond laser pulse. The bubbles are sufficiently spherical that the hydrostatic pressure gradient caused by gravity becomes the dominant source of asymmetry in the collapse and rebound of the bubbles (see Fig.~1(a) in Ref.~\citep{Obreschkow2011b}). To avoid this source of asymmetry the experiment is performed in micro--gravity conditions (ESA, 53rd parabolic flight campaign). Therefore, the experiment can be considered as producing the most spherical cavitation bubbles available at present.

The spherical bubble considered here has a maximal radius $\Rm=(2.786\pm0.007)\rm\,mm$ and collapses within $\Tc=(508.28\pm0.10)\rm\,\mu s$ under the driving pressure $p-\pv=(25.34\pm0.15)\rm\,kPa$. This bubble is centered inside a volume $(178\times178\times150)\rm\,mm^3$ of demineralized water at $(26\pm0.5)^\circ\rm C$. The bubble radius $R_{\rm obs}(T)$ is measured at sub--micron precision from a movie obtained with a high--speed camera, operating at inter--frame spacings of $10\rm~\mu s$ with exposure times of $370\rm~ns$. The high-speed movie and complementary data are available online \footnote{The movie and complementary data are published at http://bubbles.epfl.ch/data as ``cavity00096.zip''.}.

Four selected time--frames of the collapsing bubble are shown in Fig.~\ref{fig_models}(a). The evolution of the observed normalized radius $r_{\rm obs}\equiv R_{\rm obs}/\Rm$ is plotted in Fig.~\ref{fig_models}(b). We find that $r_{\rm obs}(t)$ closely follows the Rayleigh solution $r(t)$, as emphasized by the residual $r_{\rm obs}(t)-r(t)$, shown in Fig.~\ref{fig_models}(c). At no time does $r_{\rm obs}(t)$ differ by more than $10^{-3}$ from $r(t)$. Therefore, considering the different approximations $r_\ast(t)$ and $r_n(t)$ with $n\leq5$, we see that only the accuracy of $r_\ast(t)$ is sufficient to compare the experimental data against the Rayleigh model. Such a comparison makes it possible, for instance, to efficiently analyze the remaining oscillatory residual $r_{\rm obs}(t)-r_\ast(t)$, which may be explained in terms of surface tension and viscosity \citep{Plesset1977}, liquid compressibility \citep{Prosperetti1987} and thermal effects \citep{Dergarabedian1953}. Basic estimates of these effects unveil that the excellent match between $r_{\rm obs}(t)$ and $r(t)$ is partially due to a compensation of surface tension, which accelerates the collapse, by compressibility and non--condensable gases.


\section{Conclusions}\label{section_discussion}
In summary, we have developed analytical approximations to the solution $r(t)$ of the RE, which preserve the time--symmetry, the boundary conditions at $t=0$ (to arbitrary order), the branch point singularities at $t=\pm1$, and the asymptotic behavior of $\dot{r}(t)$ at $t=\pm1$. Despite their elementary forms, the zeroth-order approximation $r_0(t)$ yields a maximal error $\epsilon(r_0)$ of only about 0.01, whereas the best approximation $r_\ast(t)$, expressed in terms of the polylogarithm, reduces this error to below $10^{-5}$. Moreover, approximations $r_n(t)$ with any smaller accuracy can be systematically constructed. For example, if 0.1\% accuracy is desired then $r_2(t)\approx(1-t^2)^{2/5}(1-0.01832t^2-0.00399t^4)$ suffices. A comparison of these approximations against state-of-the-art measurements of highly spherical cavitation bubbles revealed that the residuals $r_\ast(t)-r(t)$ are more than 100 times smaller than the observed residuals $r_{\rm obs}(t)-r(t)$, shown in Fig.~\ref{fig_models}(c). Thus the approximation $r_\ast(t)$ is by far sufficient for all practical purposes.


This work was supported by the Swiss NSF (200020-116641).


\end{document}